# Pbca-type $In_2O_3$: the lost pressure-induced post-corundum phase

Braulio García-Domene,[1] Juan A. Sans,[2]* Óscar Gomis,[3] Francisco J. Manjón,[2] Henry M. Ortiz,[2] Daniel Errandonea,[1] David Santamaría-Pérez,[1] Domingo Martínez-García,[1] Rosario Vilaplana,[3] André L. J. Pereira,[2] Ángel Morales-García,[4] Plácida Rodríguez-Hernández,[5] Alfonso Muñoz,[5] Catalin Popescu,[6] and Alfredo Segura[1]

[1] ICMUV-Departamento de Física Aplicada, MALTA Consolider Team, Universitat de València, 46100 Burjassot (València), Spain
[2] Instituto de Diseño para la Fabricación y Producción Automatizada, MALTA Consolider Team, Universitat Politècnica de València, 46022 València (Spain)
[3] Centro de Tecnologías Físicas, MALTA Consolider Team, Universitat Politècnica de València, 46022 València (Spain)
[4] Departamento de Física Fundamental II, MALTA Consolider Team, Universidad de La Laguna, 38205 La Laguna, Tenerife (Spain)
[5] Departamento de Química-Física I, MALTA Consolider Team, Universidad Complutense de Madrid, 28040 Madrid (Spain)
[6] ALBA-CELLS, 08290 Cerdanyola, Barcelona (Spain)

*Corresponding author: juasant2@upv.es

# Abstract

Contradictory results of high-pressure studies in cubic bixbyite-type indium oxide (c-$In_2O_3$) at room temperature (RT) have motivated us to perform high-pressure powder x-ray diffraction and Raman scattering measurements in this material. On increasing pressure c-$In_2O_3$ undergoes a transition to the $Rh_2O_3$-II structure. On decreasing pressure $Rh_2O_3$-II-type $In_2O_3$ undergoes a transition to a previously unknown phase which is isostructural to $Rh_2O_3$-III. On further decrease of pressure, another phase transition to corundum-type $In_2O_3$, which is metastable at room conditions, is observed. Recompression of metastable corundum-type $In_2O_3$ shows that the $Rh_2O_3$-III phase is the post-corundum phase. Our results are supported by theoretical ab initio calculations which show that the $Rh_2O_3$-III phase could be present in other sesquioxides, thus leading to a revision of the pressure-temperature phase diagrams of sesquioxides.

**I. Introduction**

Indium oxide ($In_2O_3$) is a sesquioxide crystallizing in the cubic bixbyite-type (c-$In_2O_3$) structure (space group (SG) Ia-3, No. 206, Z=16) at room conditions and is employed as a gas sensor, [1] in thermoelectrics,[2] and specially, when doped with Sn or F, as a transparent conductive oxide in many optoelectronic applications including solar cells, light-emitting diodes, and liquid-crystal displays.[3] For many years $In_2O_3$ was also known in the rhombohedral corundum-type (rh-$In_2O_3$) structure (SG R-3c, No. 167, Z=6) which is metastable at ambient conditions and was obtained after application of extreme conditions of pressure and temperature.[4] The synthesis of rh-$In_2O_3$ at room conditions[5] has allowed the characterization of its properties for its use in industrial applications.[6] Therefore, the discovery of new phases of $In_2O_3$ is of fundamental and technological interest.

The finding of the post-perovskite phase in $MgSiO_3$[7] led to a major revision of the pressure-temperature phase diagrams in many $ABO_3$ compounds including sesquioxides (A=B). This led to the prediction of the existence of the post-perovskite phase in $In_2O_3$,[8] so considerable experimental efforts were done in order to find this phase at high pressures.[9,10] However, experiments showed that the post-perovskite phase in $In_2O_3$ is not stable at high pressures. Instead, two new high-pressure orthorhombic phases were discovered: $Rh_2O_3$-II (SG Pbcn, No. 60, Z=4), and α-$Gd_2S_3$ phase (SG Pnma, No. 62, Z=4).[9,10] Hereafter, $Rh_2O_3$-II and α-$Gd_2S_3$ phases of $In_2O_3$ will be named o1-$In_2O_3$ and o2-$In_2O_3$, respectively.

Most of the discoveries of new phases of $In_2O_3$ were done at high pressures and high temperatures because of the presence of kinetic effects which prevented the phase transitions at RT.[9-11] In this respect, several contradictory RT high-pressure studies of $In_2O_3$ have been recently reported up to 30 GPa. Two recent works reported a phase transition from c-$In_2O_3$ to rh-$In_2O_3$ above 12 and 25 GPa in bulk and nanocrystalline In2O3,[12a,12b] while a third recent work did not find any sign of phase transition up to 30 GPa neither in bulk nor in nanocrystalline c-$In_2O_3$.[12c] This last result was in good agreement with other previous works.[10b]

## II. Experimental Section

Commercial c-$In_2O_3$ powders of high purity (99.99%) from Sigma-Aldrich Inc. were used. rh-$In_2O_3$ powders were obtained on downstroke from cubic powders pressurized to 50 GPa at RT. High-pressure angle-dispersive powder X-ray diffraction (XRD) experiments at RT up to 50 GPa for c-$In_2O_3$ and up to 25 GPa for rh-$In_2O_3$ were conducted in a membrane-type diamond anvil cell (DAC) with a culet of 350 μm in diameter. Either He gas or a 16:3.1 methanol-ethanol-water mixture (see Figure 4) were used as quasihydrostatic pressure-transmitting media. EOS of copper was used as pressure callibrant.[13] Experiments were performed at the MSPD beamline of the ALBA synchrotron[14] with an incident monochromatic wavelength of 0.4246 Å focused to 20 x 20 μm. Images in a 2θ range up to 15° were collected with a SX165 CCD placed at 270 mm from sample. Intensity diffraction profiles as a function of 2θ were obtained with the Fit2D software. Lattice parameters of powder XRD patterns and Rietveld refinements were obtained with GSAS[15] and Powdercell [16] software, respectively.

Raman scattering (RS) measurements at RT excited either with 532.0 or 632.8 nm laser lines and laser power below 10 mW were performed in backscattering geometry using a LabRam HR UV spectrometer in combination with a thermoelectrically-cooled multichannel CCD detector (resolution below 2 $cm^{-1}$). High-pressure RS measurements of c-$In_2O_3$ up to 50 GPa and of rh-$In_2O_3$ up to 33 GPa, respectively, were performed in a 280 mm culet DAC. A 16:3:1 methanol-ethanol-water mixture was used as pressure-transmitting medium and a few ruby balls evenly distributed in the pressure chamber were employed as a pressure sensor. [17]

## III. Calculations details

*Ab initio* total-energy calculations were performed within the density functional theory using the planewave method and the pseudopotential theory with the Vienna *ab initio* simulation package (VASP).[18] We have used the projector-augmented wave scheme (PAW)[19,20] implemented in this package to take into account the full nodal character of the all-electron charge density in the core region. Basis set including plane waves up to an energy cutoff of 520 eV were used in order to achieve highly converged results and accurate description of the electronic and dynamical properties. The description of the exchange-correlation energy was described with the generalized gradient approximation (GGA) with the PBEsol prescription.[21] A dense special k-points sampling for the Brillouin Zone (BZ) integration was performed in order to obtain very well converged energies and forces. At each selected volume, the structures were fully relaxed to their equilibrium configuration through the calculation of the forces and the stress tensor. It allows to obtain the relax structures at the theoretical pressures defined by the calculated stress. In the relaxed equilibrium configuration, the forces on the atoms are less than

0.006 eV/Å, and the deviation of the stress tensor from a diagonal hydrostatic form is less than 1 kbar (0.1 GPa). The application of DFT based total-energy calculations to the study of semiconductors properties under high pressure has been reviewed,[22] showing that the phase stability, electronic, and dynamical properties of compounds under pressure are well described by DFT. In particular, enthalpy vs. pressure curves for competitive phases (Ia-3, R-3c, Pbca, and Pbcn) in $In_2O_3$ are depicted in Figure 1.

Lattice-dynamics calculations of the phonons at the zone center (Γ point) of the Brillouin zone (BZ) enable us to assign the Raman modes observed for the different phases. Furthermore, the calculations provide information about the symmetry of the modes and polarization vectors. Highly converged results on forces are required for the calculation of the dynamical matrix. We use the direct force constant approach.[23] The construction of the dynamical matrix at the Γ point of the BZ is particularly simple and involves separate calculations of the forces in which a fixed displacement from the equilibrium configuration of the atoms within the primitive unit cell is considered. Symmetry aids by reducing the number of such independent displacements, reducing the computational effort in the study of the analysed structures considered in this work. Diagonalization of the dynamical matrix provides both the frequencies of the normal modes and their polarization vectors. This allows us to identify the irreducible representation and the character of the phonons modes at the Γ point.

**IV. Results and discussions**

On upstroke to 50 GPa, powder XRD patterns of c-$In_2O_3$ (Figure 2a) can be clearly indexed with the Miller indexes of the Bragg reflections for c-$In_2O_3$ (see Rietveld refinement of the XRD pattern at 1.7 GPa with $R_{wp}$=3.5% in Figure 2) below 35 GPa. Above this pressure there is an onset of a high-pressure phase transition which is completed at 46 GPa. Above this pressure the XRD pattern fully corresponds to o1-$In_2O_3$ (see XRD pattern refined with Le Bail analysis at 50.1 GPa in Figure 3). The onset pressure for the transition to o1-$In_2O_3$ (35 GPa with helium) decreases under less hydrostatic conditions (31 GPa with 16:3:1 methanol-ethanol-water mixture) as shown in Figure 4. Therefore, our results clearly indicate that the post-cubic phase of $In_2O_3$ at RT under quasihydrostatic conditions is the orthorhombic $Rh_2O_3$-II structure and not the corundum-type phase, as it was previously.

On downstroke from 50 GPa, powder XRD patterns (Figure 2b) can be clearly indexed with Bragg reflections of o1-$In_2O_3$ down to 20 GPa. Below this pressure a transition to a previously unknown phase occurs. The XRD pattern at 17.1 GPa corresponds mainly to the new phase with some remnant peaks of o1-$In_2O_3$ while at 12.1 GPa the XRD pattern corresponds entirely to the new phase. On decreasing pressure below 4.2 GPa another phase transition occurs and the new peaks can be indexed with rh-$In_2O_3$ at room pressure (see Rietveld refinement in the released sample with $R_{wp}$=2.3% in Figure 3).

In order to look for candidate structures of the new phase we resorted to *ab initio* total-energy calculations of a number of structures already found or suggested to occur in $ABO_3$ compounds. In particular, we focused in structures which could be intermediate between the R-3c and the Pbcn phases, like the monoclinic P2/c, which was suggested to constitute a diffusionless pathway between both R3-c and Pbcn structures.[10c,24] Our calculations (Figure 1) show that the most competitive phase with the Ia-3, R3-c, and Pbcn phases is not the P2/c structure, whose enthalpy vs. pressure curve mimics that of the R-3c phase at all pressures (not shown), but the orthorhombic $Rh_2O_3$-III phase (SG Pbca, No.61, Z=8),[25] hereafter named o3-$In_2O_3$.

A good Rietveld refinement ($R_{wp}$= 1.1%) of the XRD pattern at 12.1 GPa on downstroke with the Pbca phase is shown in Figure 3. The structure obtained by this refinement was deposited into the FIZ-Karlsruhe database (CSD-number: 426846). Table I shows the experimental values of the structural parameters for o3-$In_2O_3$ at 12.1 GPa. Several differences can be noted between the XRD pattern of o3-$In_2O_3$ and those of o1-$In_2O_3$ and rh-$In_2O_3$. o3-$In_2O_3$ shows a peak below 8º that it is not present in o1-$In_2O_3$. On the other hand, one broad peak is observed above 9º in

o3-In$_2$O$_3$, corresponding to several close reflections, while two narrow peaks are observed in rh-In$_2$O$_3$.

The unit cell of o3-In$_2$O$_3$ (see Figure 5) contains 8 formula units and all 5 atoms (two In and three O) are in 8c Wyckoff positions. The structure can be understood as a distorted bixbyite-like structure where In is six-fold coordinated as well as in c-In$_2$O$_3$, rh-In$_2$O$_3$, and o1-In$_2$O$_3$. Thus, In$_2$O$_3$ could be one of the few compounds having four polymorphs with the same cation coordination. However, they have different densities since their unit cell volumes decrease in the sequence Ia-3 > R-3c > Pbca > Pbcn, as can be deduced from their equations of state (EOS) plotted in Figure 3 of the supplementary material. The corresponding densities for the Ia-3, R-3c, Pbca, and Pbcn phases are 7.179(1) g/cm$^3$, 7.255(1) g/cm$^3$, 7.449(1) g/cm$^3$ and 7.536(1) g/cm$^3$, respectively. The experimental and theoretical parameters of the EOS of the polymorphs of In$_2$O$_3$ can be found in Table II.

In order to get a deeper understanding of the sequence of pressure-induced phase transitions observed in In$_2$O$_3$ we have pressurized rh-In$_2$O$_3$ samples recovered from pressurizing c-In$_2$O$_3$ samples to 50 GPa. Figure 4 shows XRD and RS measurements of mainly rh-In$_2$O$_3$ up to 25 and 33 GPa, respectively. All XRD patterns up to 25 GPa can be indexed by a mixture of rh-In$_2$O$_3$ and o3-In$_2$O$_3$, being rh-In$_2$O$_3$ the dominant phase below 15 GPa and o3-In$_2$O$_3$ the dominant phase above this pressure. On decreasing pressure the recovered sample at 1 atm was completely indexed with rh-In$_2$O$_3$. This sample was recovered and studied by RS at high pressures (see Figure 6b). Only Raman peaks of rh-In$_2$O$_3$ are present below 12.4 GPa. Above this pressure, new peaks appear revealing a gradual phase transition to a new phase with coexistence of the two phases up to 26 GPa. The new Raman peaks appearing above 12.4 GPa are the same already attributed to o3-In$_2$O$_3$ on downstroke from 50 GPa (see Supporting Information). Between 26 GPa and 33 GPa only Raman modes of o3-In$_2$O$_3$ are observed. Finally, rh-In$_2$O$_3$ is obtained metastably on decreasing pressure to 1 atm. The good agreement between XRD and RS results confirm that the Pbca phase is the post-corundum phase in In$_2$O$_3$.

Scheme 1 summarizes the sequence of pressure-induced phase transitions in In$_2$O$_3$ at RT up to 50 GPa. In this work, the four polymorphs of In$_2$O$_3$ were obtained starting from the cubic polymorph by compression and decompression at RT without high temperature treatment. It remains to be proved whether the Pbca phase in In$_2$O$_3$ could be metastable obtained at room conditions for its use in new applications. We hope the results here reported will trigger such studies.

We want to finish by highlighting that the discovery of the Pbca phase of In$_2$O$_3$ as a high-pressure post-corundum phase has far-reaching implications for the understanding of the pressure-temperature diagram of sesquioxides. The Pbca phase was previously found as a high-temperature phase of Rh$_2$O$_3$ (Rh$_2$O$_3$-III)[25] and also in Mn$_2$O$_3$.[26] This phase is so poorly known in Mn$_2$O$_3$ that it has been recently postulated to be its low-temperature phase since Mn$_2$O$_3$ usually crystallizes in the cubic phase.[27] Moreover, a new mineral named panguite, mainly composed of Ti$_2$O$_3$ with Pbca phase has been recently found in the Allende meteorite.[28] These results show that the Pbca phase is one of the less known and understood phases of sesquioxides and that it could be a high-pressure phase in other sesquioxides. In fact, our calculations (not shown here) indicate that this phase could be present in other corundum-type sesquioxides, like Al$_2$O$_3$, Ga$_2$O$_3$, V$_2$O$_3$, Cr$_2$O$_3$, and Fe$_2$O$_3$, so detailed experimental studies on upstroke and downstroke should be performed to confirm its presence in other sesquioxides.

**V. Summary**

In summary, our XRD measurements provide evidence for four different polymorphs of In$_2$O$_3$ at RT in the pressure range to 50 GPa, being the Pbca-type (o3-In$_2$O$_3$) structure here reported for the first time. All the phase transitions determined by powder XRD have been also confirmed by means of Raman scattering (RS) measurements on upstroke and downstroke and are supported by our *ab initio* total-energy and lattice-dynamics calculations.

**VI. Acknowledgements.**

Financial support by the Spanish MEC under Grant No. MAT2010-21270-C04-01/03/04, by MALTA Consolider Ingenio 2010 project (CSD2007-00045) and by Generalitat Valenciana (GVA-ACOMP-2013-012). Red Española de Supercomputación (RES) and ALBA Synchrotron Light Source are also acknowledged. BGD and JAS acknowledge financial support through the FPI program and Juan de la Cierva fellowship. We also thank J. L. Jordá for fruitful discussions.

# TABLES

| | a (Å) | b (Å) | c (Å) | Atom | Wyckoff site | x | y | z |
|---|---|---|---|---|---|---|---|---|
| Experimental | 5.3934(14) | 5.5002(15) | 15.589(4) | In1 | 8c | 1.006(5) | 1.753(4) | 0.9528(5) |
| | | | | In2 | 8c | 0.006(5) | 0.754(5) | 0.6823(5) |
| | | | | O1 | 8c | 0.3595 | 0.8772 | 0.7016 |
| | | | | O2 | 8c | 0.2066 | 0.5233 | 0.8752 |
| | | | | O3 | 8c | 0.1465 | 0.6581 | 0.5526 |

**Table I.** Experimental (from Rietveld refinement) values of the lattice parameters and atomic positions obtained for the Pbca phase of $In_2O_3$ at 12.1 GPa. Due to the large number of variables, O atomic positions were assumed to be fixed while In atomic positions were refined.

| Polymorph | Origin | $V_0$ (Å$^3$) | $B_0$ (GPa) | $B_0$' |
|---|---|---|---|---|
| Ia-3 | Experimental | 64.28(13) | 184(10) | 4.0 |
| | Theoretical | 65.72(2) | 169.4(12) | 4.0 |
| Pbca | Experimental | 61.9(3) | 189(40) | 4.0 |
| | Theoretical | 62.9(1) | 169.6(14) | 4.0 |
| Pbcn | Experimental | 61.2(7) | 194(10) | 4.0 |
| | Theoretical | 61.8(2) | 185.4(12) | 4.0 |

**Table II.** Experimental and theoretical values of the unit-cell volume ($V_0$), bulk modulus ($B_0$), and pressure derivative ($B_0$') at room pressure for the three polymorphs (Ia-3, Pbca and Pbcn) of $In_2O_3$. Data for the R-3c phase are not given due to the lack of enough experimental data. Both experimental and theoretical data have been fitted to a second-order Birch-Murnaghan equation of state ($B_0$' fixed to 4).

**FIGURES**

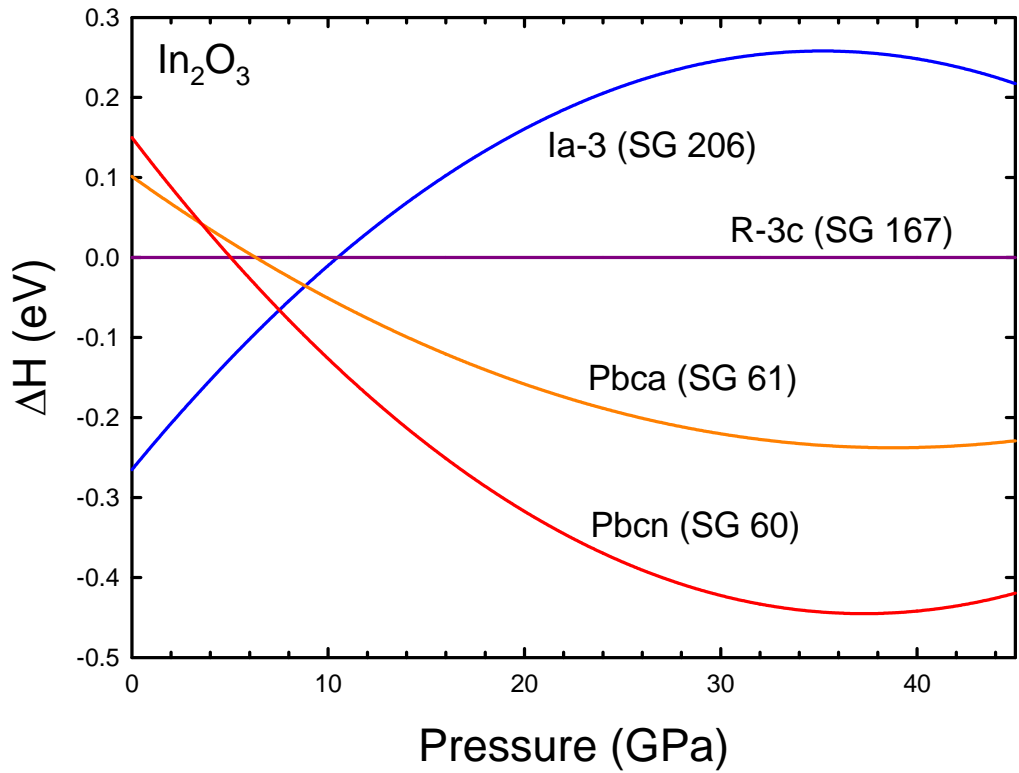

**Figure 1.** Calculated enthalpy-pressure diagram for indium oxide polymorphs with respect to the corundum-type rhombohedral R-3c phase (vilolet): cubic Ia-3 (blue), orthorhombic Pbca (orange), and orthorhombic Pbcn (red).

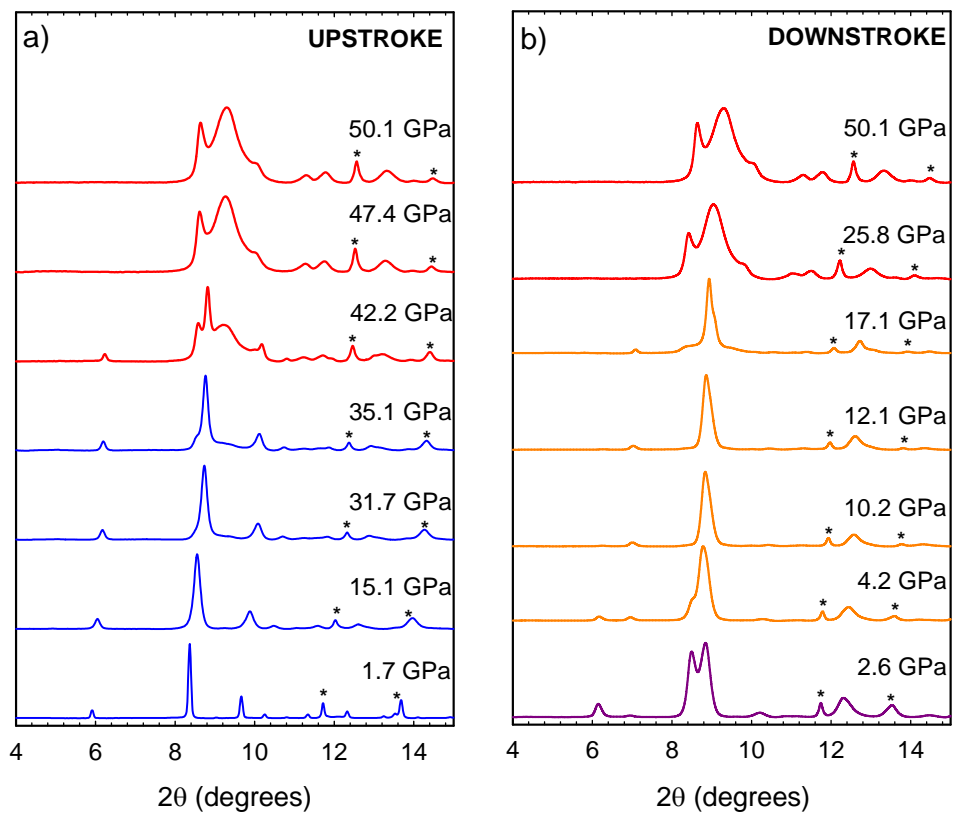

**Figure 2.** XRD powder patterns of c-In$_2$O$_3$ at RT and different pressures on upstroke up to 50 GPa (a) and on downstroke down to 2.6 GPa (b). Colours indicate the XRD patterns with mainly Ia-3 (blue), Pbcn (red), Pbca (orange), and R-3c (violet) phases. Asterisks denote the diffraction peaks of Cu.

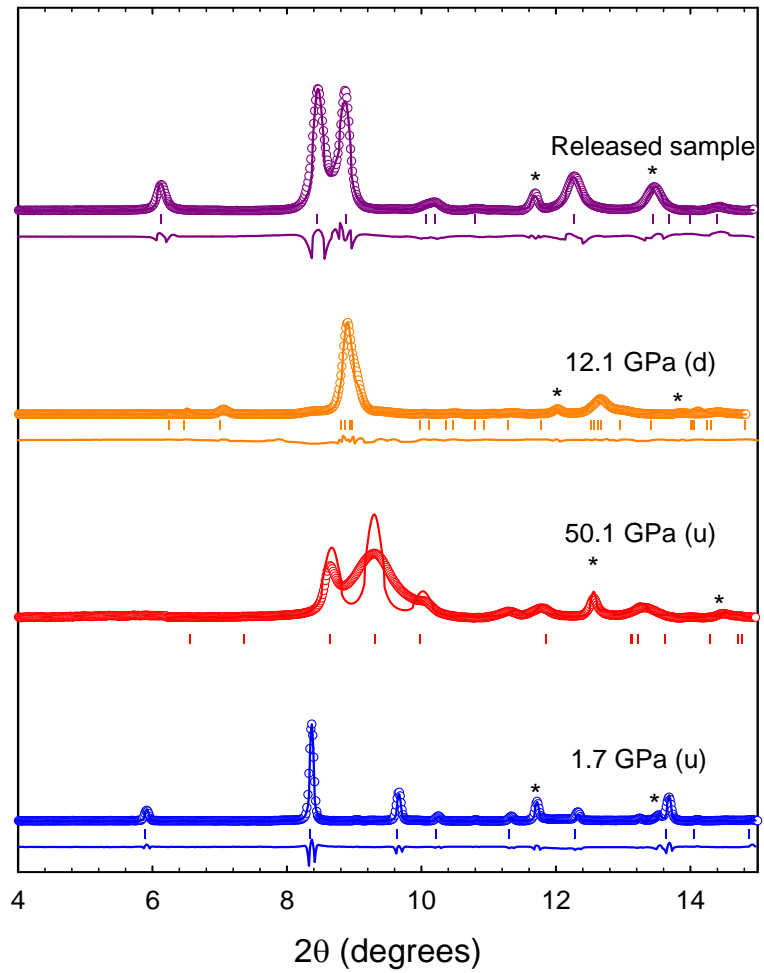

**Figure 3.** Experimental XRD patterns on upstroke (u) at 1.7 GPa with the Ia-3 structure and at 50.1 GPa with the Pbcn structure and on downstroke (d) at 12.5 GPa with the Pbca structure and at 1 atm with the R-3c structure. Residuals of the Rietveld refinements are plotted below the experimental (circles) and fitted (lines) XRD profiles (except for Pbcn phase, where a LeBail refinement has been performed).

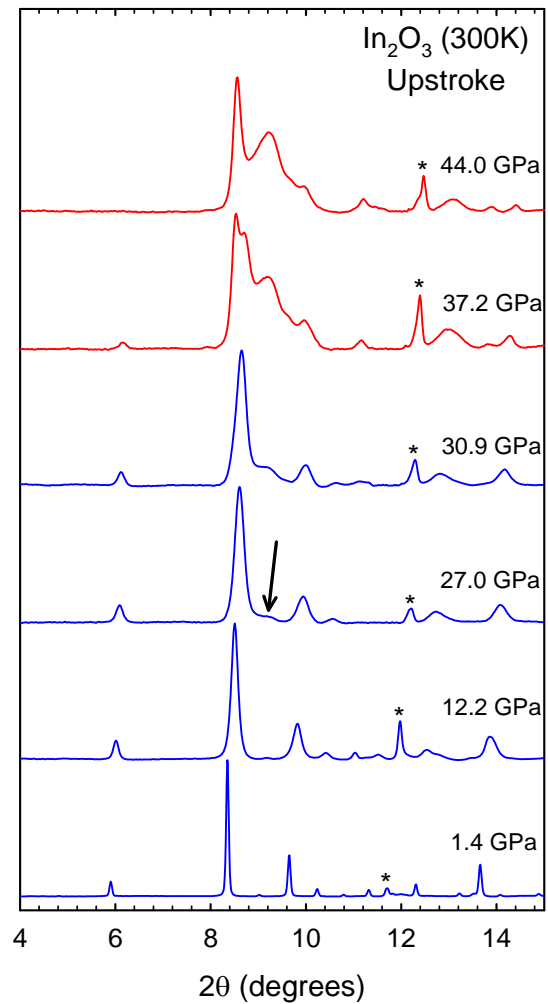

**Figure 4**. Selection of XRD powder patterns of $In_2O_3$ at room temperature and pressures up to 44 GPa. A mixture of 16:3:1 methanol-ethanol-water was used as pressure-transmitting medium. Asterisks denote the diffraction maxima of copper at each pressure. The arrow indicates the appearance of the diffraction peaks of the high-pressure phase and thus the onset of the pressure-induced phase transition.

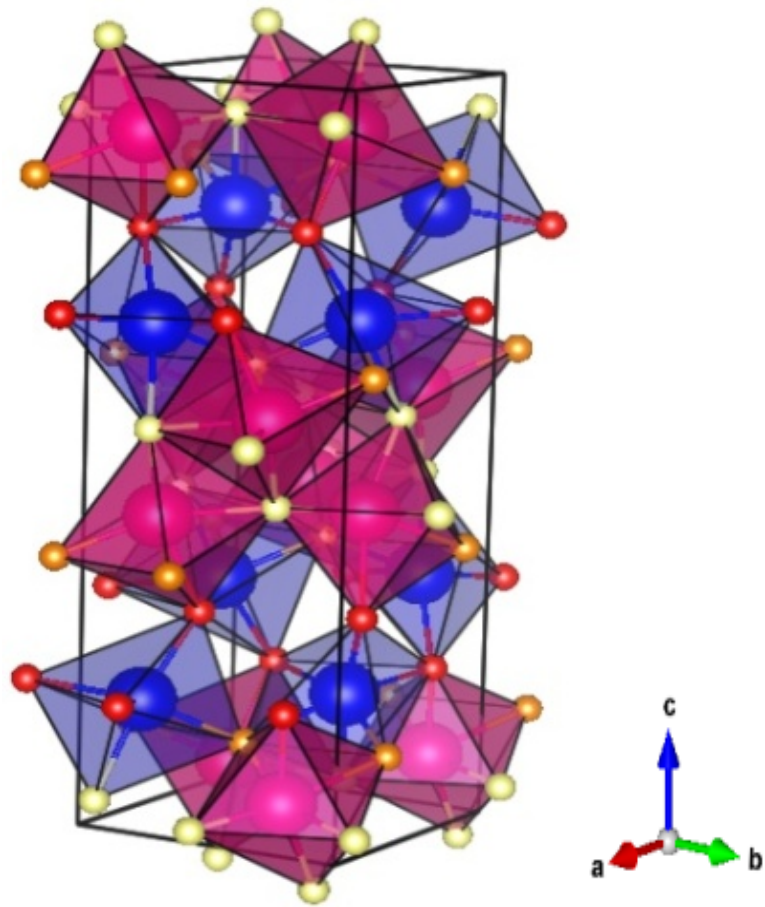

**Figure 5**. Pbca-type $In_2O_3$. Blue and pink atoms are the two inequivalent In atoms, while red, yellow, and orange atoms are the three inequivalent O atoms.

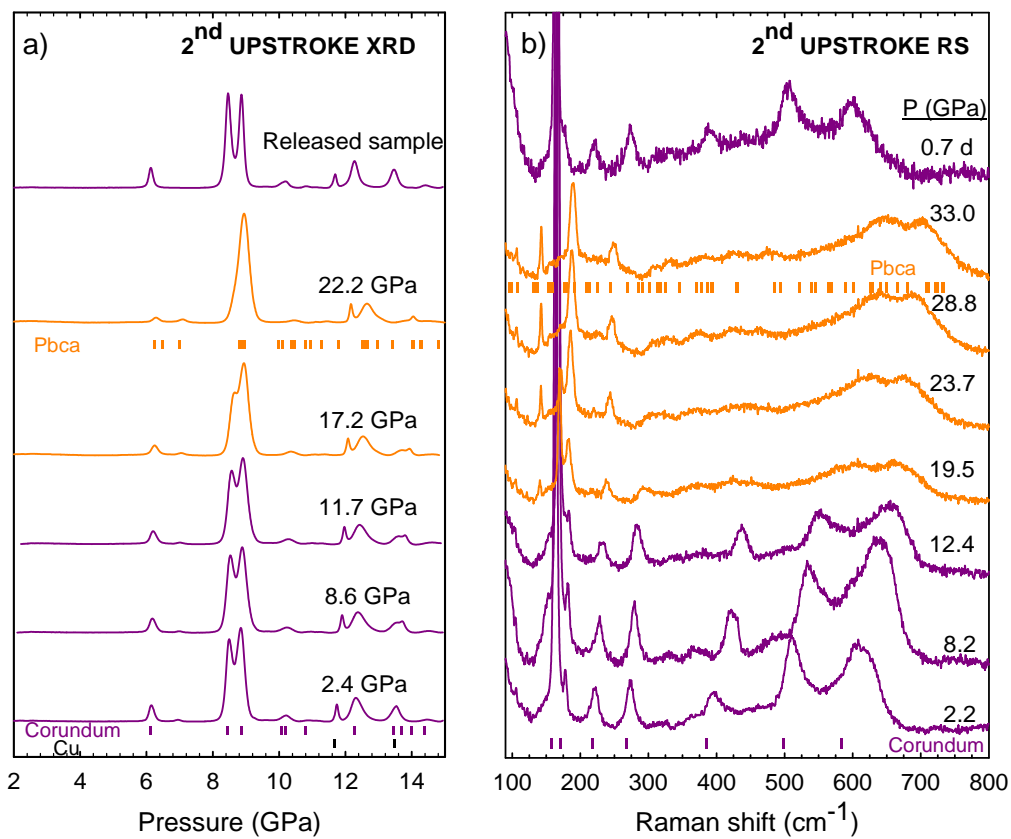

**Figure 6.** Selected XRD patterns up to 25 GPa (a) and RS spectra up to 33 GPa (b) of rh-$In_2O_3$. Spectra mainly corresponding to R-3c and Pbca, phases are represented in violet and orange, respectively. Ticks at the bottom of XRD (RS) patterns of 2.4 (2.2) and 22.2 (33.0) GPa mark the positions of the diffraction (Raman) peaks of the dominating R-3c and Pbca phases at each pressure, respectively.

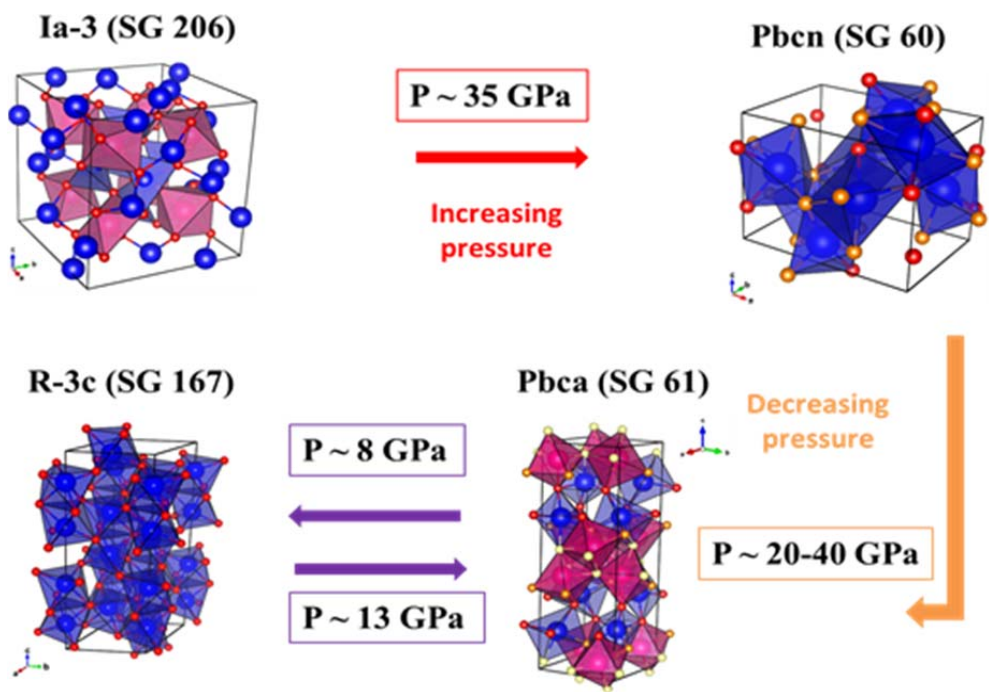

**Scheme 1.** Sequence of pressure-induced phase transitions in $In_2O_3$ up to 50 GPa. Atoms in different Wyckoff sites are depicted by different colours: In (blue and pink) and O (red, yellow and orange).